\documentclass[12pt]{article}
\usepackage{subcaption}
\usepackage{mathtools}
\usepackage{jheppub}

\usepackage{parskip}
\usepackage{xcolor}
\usepackage{colortbl}
\usepackage{tikz}
\usepackage{tikz-cd}
\usepackage{blkarray}
\usepackage{bm}

\newcommand{\bra}[1]{\langle{#1}|}
\newcommand{\ket}[1]{|{#1}\rangle}

\newcommand{\kket}[1]{\Vert{#1}\rangle\!\rangle}
\newcommand{\KT}{\text{KT}}

\newcommand{\apeven}{e5f6ff}
\newcommand{\apodd}{fff9d9}
\newcommand{\peven}{ffecf9}
\newcommand{\podd}{e7ffcd}
\newcommand{\pic}[4]{\begin{array}{r|ll} & \text{even} & \text{odd} \\ \hline \text{AP} & \cellcolor[HTML]{\apeven} #1 & \cellcolor[HTML]{\apodd} #2 \\ \text{P} & \cellcolor[HTML]{\peven} #3 & \cellcolor[HTML]{\podd} #4 \end{array}}
\newcommand{\colmath}[2]{\text{\colorbox[HTML]{#1}{$#2$}}}

\newcommand{\spinarg}[3]{\begin{tikzpicture}[scale=0.45, baseline=0] \draw (0,0) rectangle (1,1); \node[anchor=north, inner sep=0mm] at (0.5,-0.2) {\scriptsize{#2}}; \node[anchor=east, inner sep=0mm] at (-0.1,0.5) {\scriptsize{#1}}; #3 \end{tikzpicture}}
\newcommand{\spin}[2]{\spinarg{#1}{#2}{}}
\newcommand{\spinh}[2]{\spinarg{#1}{#2}{\draw[densely dashed] (0,0.5) -- (1,0.5);}}
\newcommand{\spinv}[2]{\spinarg{#1}{#2}{\draw[densely dashed] (0.5,0) -- (0.5,1);}}
\newcommand{\spind}[2]{\spinarg{#1}{#2}{\draw[densely dashed] (0,0) -- (1,1);}}
\newcommand{\sspin}[2]{\scalebox{0.8}{\spin{#1}{#2}\,}}
\newcommand{\sspinh}[2]{\scalebox{0.8}{\spinh{#1}{#2}\,}}
\newcommand{\sspinv}[2]{\scalebox{0.8}{\spinv{#1}{#2}\,}}

\makeatletter
\newcommand{\xnoarrow}[2][]{\ext@arrow 3359\noarrowfill@{#1}{#2}}
\def\noarrowfill@{\arrowfill@-\relbar{-\vphantom{\rightarrow}}}
\makeatother

\title{Boundary States and Anomalous Symmetries of Fermionic Minimal Models}

\author{Philip Boyle Smith}

\affiliation{
Department of Applied Mathematics and Theoretical Physics, \\
University of Cambridge, Cambridge, CB3 OWA, UK
}

\emailAdd{pb594@damtp.cam.ac.uk}

\abstract{The fermionic minimal models are a recently-introduced family of two-dimensional spin conformal field theories. We determine all of their conformal boundary states and potentially anomalous $\mathbb{Z}_2$ global symmetries. The latter task hinges upon on a conjecture about $\mathfrak{su}(2)$ affine parities generalising an earlier result known to have an interpretation in terms of Fermat curves. Our results indicate a close connection between several properties of the models, including the matching of the sizes of the SPT classes of boundary states, the existence of anomalous $\mathbb{Z}_2$ symmetries, and the vanishing of the Ramond-Ramond sector, for which we provide an explanation.}

\begin{document}

\maketitle

\newpage

\section{Introduction} \label{sec:int}

The fermionic minimal models are a family of two-dimensional conformal field theories recently introduced in \cite{watts, tachikawa, kulp}. They are close relatives of the bosonic minimal models that have been known for a long time \cite{bpz, zuber, kato, ruelle, minimalstates}, but differ in one key respect: instead of depending solely on a Riemann surface with metric, they also depend on a choice of spin structure.

The fact that these theories depend upon a spin structure opens up the possibility of new fermionic phenomena that are not seen in the ordinary bosonic minimal models. For a start, there can be an issue with their boundary conditions. While in the bosonic minimal models, there exists a complete set of boundary conditions such that when one places the theory on a spatial interval and imposes different boundary conditions at the two ends, one always obtains a physically sensible theory \cite{minimalstates}, in fermionic theories this no longer need be the case. Instead, one may find that for certain choices of boundary conditions, the spectrum on an interval includes an unpaired Majorana zero mode, rendering the theory inconsistent \cite{bcft}. As we review later in this introduction, the fact this issue arises only for fermionic theories is closely tied to the classification of SPT phases in two dimensions, where it is reflected in the mathematical fact that $\Omega_2^\text{SO}(\text{pt}) = 0$ is trivial while $\Omega_2^\text{Spin}(\text{pt}) = \mathbb{Z}_2$ is not \cite{kapustin1, kapustin2}.

Second, there is also the possibility of supporting symmetries with nontrivial 't Hooft anomalies. For the bosonic minimal models, this is not an option: all global symmetries are necessarily non-anomalous \cite{cftrelativeanom}. But for fermionic theories, it is a possibility. We will limit ourselves to $\mathbb{Z}_2$ global symmetries. In fermionic theories, such symmetries can carry a mod-8 valued anomaly. As was explained in \cite{ryuzhang, qi}, this anomaly is related to SPT phases in three dimensions, and the fact it arises only for fermionic theories is encoded in the facts $\Omega_3^\text{SO}(B\mathbb{Z}_2) = 0$ while $\Omega_3^\text{Spin}(B\mathbb{Z}_2) = \mathbb{Z}_8$.

The above two phenomena are easy to exhibit for the simplest fermionic minimal model, the Majorana fermion. Here, one can simply use free-field techniques to explicitly show everything one could possibly want to show. But for the other fermionic minimal models, which are all interacting, things are not so transparent. Our goal is to extend the analysis of boundary conditions and anomalous symmetries to these remaining fermionic minimal models. We would like to see which features of the Majorana fermion generalise to the family as a whole, and which are artefacts of a free theory. Therefore, to set the scene, it will be useful to first review the basic facts about the Majorana fermion we want to generalise. A summary of our results follows, and after that the organisation of the paper.

\subsection{A Simple Illustration} \label{sec:int:maj}

We begin with a motivating discussion about SPT phases. A Majorana fermion of mass $m$ in $d=1+1$ dimensions has action
\[ S = \frac{i}{2} \int \! dt dx \, \chi_+ \partial_+ \chi_+ + \chi_- \partial_- \chi_- + m \chi_+ \chi_- \]
where $\partial_\pm = \partial_t \pm \partial_x$. When $m \neq 0$ the theory is gapped, and the two possible gapped phases with $m > 0$ and $m < 0$ are examples of two distinct SPT phases. To see the distinction, we promote $m \rightarrow m(x)$ and consider a domain-wall profile for $m(x)$, say one with $m(x = \pm \infty) = \pm M$ for some unimportant constant $M > 0$. As was famously shown by Jackiw and Rebbi \cite{jr}, the spectrum of the theory then includes a quantum-mechanical zero mode localised on the domain wall, obeying
\[ \chi = \chi^\dagger \qquad \{\chi, \chi\} = 1 \]
This system exhibits an anomaly under fermion parity $(-1)^F$. (One way to see this is that $\chi$ acquires a nonzero VEV on a periodic temporal circle.) The appearance of this anomalous degree of freedom on the interface signals that the $m > 0$ and $m < 0$ phases are distinct SPT phases, as promised \cite{wittentalk}.

As stressed in \cite{ryu, han, shen1, shen2}, this story has an alternative guise in the language of boundary conditions. Recall that for a massless Majorana fermion there are two possible boundary conditions, which we denote as
\begin{align*}
+ \, &{:} \quad \chi_+ = +\chi_- \\
- \, &{:} \quad \chi_+ = -\chi_-
\end{align*}
Consider a mass profile $m(x)$ that interpolates from $0$ to $\pm M$. On the left side lives a massless Majorana fermion, while the right side is gapped at a scale $M$. At low energies, the massless fermion experiences this set-up as a boundary condition. Pictorially, the map from gapped phases to boundary conditions is
\begin{equation*}
\includegraphics[scale=.6]{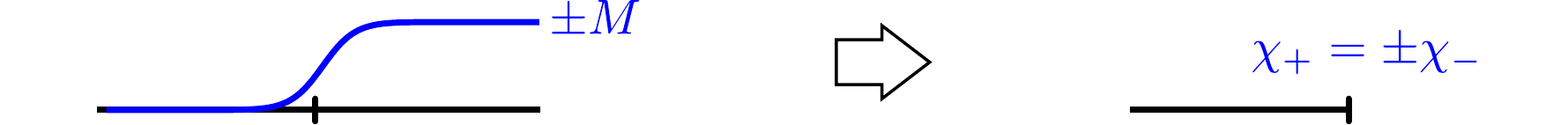}
\end{equation*}
Similarly when the gapped phase sits on the left, one obtains a left boundary condition. However, due to an annoying technicality, the outcome is now reversed:
\begin{equation}
\includegraphics[scale=.6]{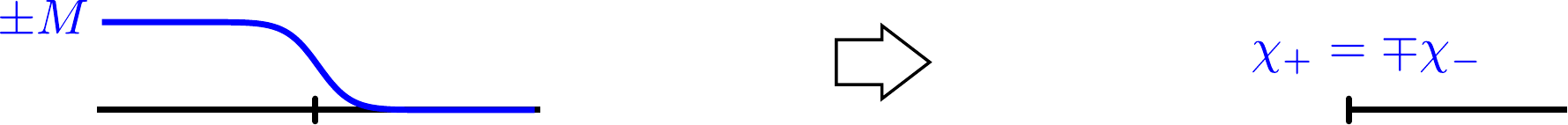} \label{eqn:sgnflip} \tag{$\star$}
\end{equation}
The analog of the previous story is now that on a spatial interval with $++$ boundary conditions at both ends, the spectrum again includes an unpaired Majorana zero mode. A lattice version of this mechanism also exists, and was the subject of \cite{kitaev}.

In conformal field theories, boundary conditions are more naturally described as boundary states. It is these that will form the main focus of the paper. To each right boundary condition, there is an associated boundary state, which lives in the NS sector of the theory:
\[ \chi_+ = \pm \chi_- \text{ on right} \quad \rightarrow \quad \ket{\pm} \]
Meanwhile, boundary conditions on the left are described by dual states. One might have thought that the correct dual state that describes boundary condition $A$ on the left is simply the dual of the state that describes boundary condition $A$ on the right. But this isn't quite right. Annoyingly, there is an extra sign flip, and the correct mapping of boundary conditions to dual states is actually
\[ \chi_+ = \pm \chi_- \text{ on left} \quad \rightarrow \quad \bra{\mp} \]
Because of the similar sign flip in \eqref{eqn:sgnflip}, however, a gapped phase $m = \pm M$ corresponds to the same boundary state $\bra{\pm}$ or $\ket{\pm}$ regardless of whether we sit at a left or a right boundary. It is for this reason that boundary states more precisely correspond to SPT phases than boundary conditions. In any case, once the boundary states are known, there is a simple algebraic procedure to calculate the partition function for states on an interval. Doing this for the boundary states $\bra{-}$ and $\ket{+}$ yields the partition function
\[ \mathcal{Z}_{-+}(\tau) = \sqrt{2} \, \chi_{1/16}(\tau) \]
where $\chi_{1/16}(\tau)$ is a Virasoro character. The most important feature of this partition function is the overall factor of $\sqrt{2}$. This signals the unpaired Majorana mode, hence the distinctness of the SPT phases underlying $\bra{-}$ and $\ket{+}$. For more details of this point of view on unpaired Majorana modes, see also \cite{bcft, flows, delmastro}.

For the second half of our story, we will be interested in the $\mathbb{Z}_2$ global symmetry known as chiral fermion parity. It acts by flipping the sign of only one of the fermions, which without loss of generality we can take to be the left-movers:
\[ \mathllap{\mathbb{Z}_2 \, {:} \quad} \chi_+ \rightarrow \chi_+ \qquad \chi_- \rightarrow -\chi_- \]
Under this symmetry, the mass parameter $m$ is odd. The symmetry therefore exchanges the two SPT phases. One can check that the symmetry also exchanges the corresponding boundary states
\[ \mathllap{\mathbb{Z}_2 \, {:} \quad} \ket{\pm} \rightarrow \ket{\mp} \]
as expected.

\newpage

Importantly, the symmetry also carries an anomaly whose strength is 1 mod 8. To see this, one places the theory on a background with a defect line for the $\mathbb{Z}_2$ symmetry. It will be sufficient for us to consider a torus. We will denote the partition function on such a background by
\[ \mathcal{Z} \Big[ \tau; \, \spinv{AP}{P} \; \Big] \]
where $\tau$ is the modular parameter describing a choice of flat metric on the torus, the diagram labels the spin structure, and the dashed line, if present, labels the defect. Under the shift $\tau \rightarrow \tau + 2$, we have the transformation law
\[ \mathcal{Z} \Big[ \tau + 2; \, \spinv{AP}{P} \; \Big] \; = \; e^{2 \pi i / 8} \; \mathcal{Z} \Big[ \tau; \, \spinv{AP}{P} \; \Big] \]
The presence of the phase $e^{2 \pi i / 8}$ indicates the anomaly. In general, it could be any eighth root of unity $e^{2 \pi i k / 8}$; the fact that for the Majorana fermion $k = 1$ indicates that the strength of the anomaly is 1 mod 8, as claimed \cite{anomalyinterplay, wittenphases}. The anomaly also manifests itself in a far more obvious way, which becomes clear if we look at the partition function. It turns out to be
\[ \mathcal{Z} \Big[ \tau; \, \spinv{AP}{P} \; \Big] \; = \; \sqrt{2} \; \overline{(\chi_0 + \chi_{1/2})(\tau)} \, \chi_{1/16}(\tau) \]
Once again we find a notorious factor of $\sqrt{2}$ in the partition function. This time it is telling us that the R sector, when frustrated by a $\mathbb{Z}_2$ symmetry defect, contains an unpaired Majorana zero mode. (This statement would also have been true of the NS sector.)

\subsection{Summary of Results} \label{sec:int:sum}

We now consider generalising these facts to the other models in the family. Each model should have a complete set of conformal boundary states. We expect that all of these states will arise from a deformation to a gapped phase. If so, then each boundary state will fall into one of two distinct classes depending on which SPT class its gapped phase sits in.

Our first main claim is that this expectation is borne out. We determine, for each fermionic minimal model, the complete list of conformal boundary states. We show how these naturally fall into two classes. When boundary conditions $a$ and $b$ are taken from the same class, we find the partition function
\[ \mathcal{Z}_{ab}(\tau) = \sum_{i \in \KT} n_{ab}^i \, \chi_i(\tau) \]
where $\KT$ denotes the Kac table, as we review in Section~\ref{sec:review}. The coefficients $n_{ab}^i$ are non-negative integers, so this is a manifestly sensible partition function. In contrast, when $a$ and $b$ are from different classes, we find
\[ \mathcal{Z}_{ab}(\tau) = \sqrt{2} \, \sum_{i \in \KT} n_{ab}^i \, \chi_i(\tau) \]
which contains a factor of $\sqrt{2}$, signalling an unpaired Majorana mode, but other than that the partition function is perfectly sensible, with the coefficients $n_{ab}^i$ again given by non-negative integers. (For a previous example of this phenomenon in a different family of models, see also \cite{bcft}.) All partition functions are explicitly presented in Section~\ref{sec:states}.

\begin{figure}
\centering
\includegraphics[scale=.9]{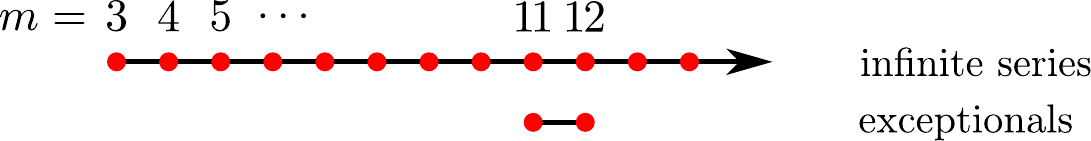}
\caption{The classification of fermionic minimal models. Both the infinite series and exceptionals are labelled by a choice of integer $m$.}
\label{fig:int:minimals}
\end{figure}

To describe these results in a little more detail, we recall that there are two kinds of fermionic minimal models: an infinite series, and two exceptionals. Both kinds are labelled by an integer $m$; for the infinite series, this integer takes the values $m \geq 3$, while for the exceptionals, it takes the values $m = 11, 12$. This situation is depicted in Figure~\ref{fig:int:minimals}. We explain our results for each kind of model in turn.

\begin{figure}
\centering
\subcaptionbox*{$m = 3$}{\includegraphics[scale=1.3]{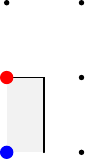}}\hfill
\subcaptionbox*{$m = 4$}{\includegraphics[scale=1.2]{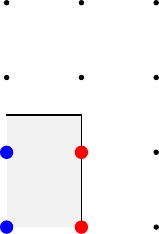}}\hfill
\subcaptionbox*{$m = 5$}{\includegraphics[scale=1.1]{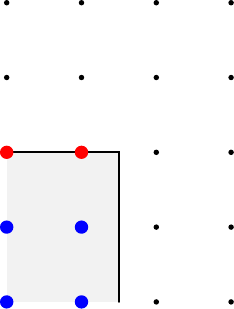}}\hfill
\subcaptionbox*{$m = 6$}{\includegraphics[scale=1]{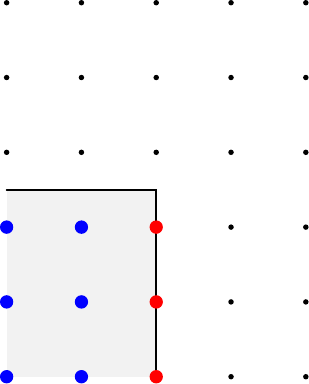}}
\caption{Boundary states for the infinite series of models are labelled by points in the bottom-left quadrant of the Kac table. The two classes are shown in blue and red. Note that for $m = 3, 4$ but no other values, the classes have equal sizes.}
\label{fig:int:states}
\end{figure}

\paragraph{Infinite series} ${}$ \\
Here the boundary states are labelled by pairs $(r, s)$ in the bottom-left quadrant of the Kac table, defined by the inequalities
\[ r \leq m/2 \quad \text{and} \quad s \leq (m+1)/2 \]
These states are shown in Figure~\ref{fig:int:states} for the first four models in the series. The states fall into two classes, with class 1 consisting of the points in the interior of the region defined by the above inequalities, and class 2 consisting of the border. It is straightforward to count the number of states in the two classes. They have sizes
\begin{align*}
\text{class 1:} &\qquad \lfloor (m - 1)^2 / 4 \rfloor \\
\text{class 2:} &\qquad \lfloor m / 2 \rfloor
\end{align*}

\paragraph{Exceptionals} ${}$ \\
In this case the labelling of the boundary states is a little more complicated and will be left to Section~\ref{sec:states}. However, the counting is straightforward. Now both classes have the same size. For the $m = 11$ exceptional, both classes have size 10, while for the $m = 12$ exceptional, both have size 12.

\begin{figure}
\centering
\vspace{1em}
\includegraphics[scale=.9]{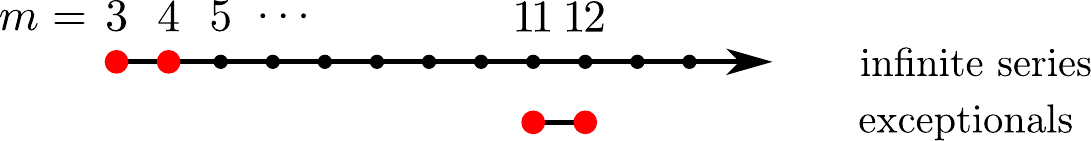}
\caption{The models with an extra $\mathbb{Z}_2$ global symmetry, shown in red. Note that these are the same models which had matching class sizes earlier, as well as the models with vanishing RR sector.}
\label{fig:int:minimalswithz2}
\end{figure}

Our second main result is a classification of which fermionic minimal models have a global $\mathbb{Z}_2$ symmetry, modulo $(-1)^F$. We find that the first two models in the infinite series and both exceptionals have a unique such symmetry, generalising the chiral fermion parity of the Majorana fermion, while the remaining models have none. This situation is depicted in Figure~\ref{fig:int:minimalswithz2}.

In the four models where the symmetry exists, we also find that the strength of the anomaly takes the same value 1 mod 8 for all models; that the twisted-sector partition functions contain a Majorana zero mode whenever there is a symmetry defect crossing the equal-time contour; and that the symmetry exchanges the two classes of boundary states. This forces them to have equal sizes, which is indeed what we found earlier. We also notice that these models are the same as the ones that have a vanishing RR sector partition function.

In view of the above results, the four special models form a close generalisation of the Majorana fermion, with all the facts we saw in Section~\ref{sec:int:maj} continuing to hold. The remaining models, on the other hand, do not form such a close generalisation, and the only fact that survives is the existence of two incompatible classes of boundary states.

Finally we mention earlier related research. The `chiral fermion parity' in the $m = 3, 4$ models is already well known. The paper \cite{watts} also obtains some of our results on boundary states for a subset of the infinite series, with a different choice of normalisation. The main novelties of this paper are: a uniform treatment of all fermionic minimal models, the normalisation, the interplay between boundary states and symmetries, and the perspective of SPT phases.

\subsection{The Plan of the Paper}

In Section~\ref{sec:review} we review a handful of facts about the minimal models and their fermionic counterparts that we will need to use, for the purposes of self-containment. In Section~\ref{sec:states}, we write down the two classes of boundary states, and explicitly compute all interval partition functions. In Section~\ref{sec:syms}, we determine all $\mathbb{Z}_2$ global symmetries, and in doing so compute their anomalies and all twisted-sector partition functions. We also compute their action on the boundary states found earlier. Finally, in Section~\ref{sec:discussion}, we give some brief arguments showing various implications between our results.

\section{Review of Fermionic Minimal Models} \label{sec:review}

In this section we review the barest essentials from the theory of minimal models that we will need to use in this paper. We will not review this material from scratch; instead, for reviews of minimal models see for example \cite{ruelle}, whose conventions we have strived to match, while for those of the fermionic ones see \cite{tachikawa, kulp}.

The fermionic minimal models are defined to be the set of all unitary spin-CFTs that are rational with respect to the Virasoro algebra. Their classification has been carried out, and the results are as in Figure~\ref{fig:int:minimals}. The classification consists of
\begin{itemize}
\item
An infinite series, with $m \geq 3$.
\item
Two exceptionals, with $m = 11, 12$.
\end{itemize}
The integer $m$ dictates the chiral data of the theory. For example, the central charge is determined via
\[ c = 1 - \frac{6}{m(m+1)} \]
Of special importance is the Kac table, which is defined as a set of pairs of integers modulo an equivalence relation,
\[ \KT \; = \; \frac{\big\{ \, (r, s) \, : \, 1 \leq r \leq m-1, \, 1 \leq s \leq m \, \big\}}{(r, s) \sim (m - r, m + 1 - s)} \]
The relevance of the Kac table is that it labels the available Virasoro characters at this central charge. We will denote them by $\chi_{r, s}(\tau)$, and their conformal dimensions by $h_{r,s}$. Sometimes, we will find it economical to compress an element of $\KT$ down to a single composite index, which we will denote by a letter such as $i,j,k,\dots$.

For the special values $m = 11, 12$, one also faces a dichotomy between the infinite series and the exceptionals. This choice affects the way the characters are combined in the partition function. We will follow the presentation of \cite{tachikawa}. We arrange the states of even/odd fermion parity on an antiperiodic/periodic\footnote{Some alternative common synonyms are AP/P, NS/R, and bounding/non-bounding.} circle into a $2 \times 2$ table. For the infinite series, this table is\footnote{Here we regard $\mathcal{Z}$ and $(-1)^\text{Arf} \mathcal{Z}$ as equivalent theories, so we only show a single table for both of them. We revisit their distinction more carefully in Section~\ref{sec:discussion}.}
\vspace{-.3em}
\begin{equation*}
\pic
{\sum_{1 \leq s \leq r < m}^{(m+1)r + ms \text{ odd}} |\chi_{r,s}|^2}
{\sum_{1 \leq s \leq r < m}^{(m+1)r + ms + m(m+1)/2 \text{ odd}} \overline{\chi_{m-r,s}} \chi_{r,s}}
{\sum_{1 \leq s \leq r < m}^{(m+1)r + ms + m(m+1)/2 \text{ even}} \overline{\chi_{m-r,s}} \chi_{r,s}}
{\sum_{1 \leq s \leq r < m}^{(m+1)r + ms \text{ even}} |\chi_{r,s}|^2}
\end{equation*}
\rule{0em}{1.2em}
For the $m = 11$ exceptional, it is
\vspace{-.3em}
\begin{equation*}
\pic
{\sum_{r = 1, \text{odd}}^9 |\chi_{r,1} + \chi_{r,7}|^2 + |\chi_{r,5} + \chi_{r,11}|^2}
{\sum_{r = 1, \text{odd}}^9 (\overline{\chi_{r,1} + \chi_{r,7}})(\chi_{r,5} + \chi_{r,11}) + \text{c.c.}}
{\sum_{r = 1, \text{odd}}^9 |\chi_{r,4} + \chi_{r,8}|^2}
{\sum_{r = 1, \text{odd}}^9 |\chi_{r,4} + \chi_{r,8}|^2}
\end{equation*}
\rule{0em}{1.2em}
while for the $m = 12$ exceptional, it is
\vspace{-.3em}
\begin{equation*}
\pic
{\sum_{s = 1, \text{odd}}^{11} |\chi_{1,s} + \chi_{7,s}|^2 + |\chi_{5,s} + \chi_{11,s}|^2}
{\sum_{s = 1, \text{odd}}^{11} (\overline{\chi_{1,s} + \chi_{7,s}})(\chi_{5,s} + \chi_{11,s}) + \text{c.c.}}
{\sum_{s = 1, \text{odd}}^{11} |\chi_{4,s} + \chi_{8,s}|^2}
{\sum_{s = 1, \text{odd}}^{11} |\chi_{4,s} + \chi_{8,s}|^2}
\end{equation*}

\section{Boundary States} \label{sec:states}

In this section we write down complete sets of boundary states for the fermionic minimal models. We compute their interval partition functions to confirm their consistency, and use them to show which states lie in which class.

First we review the general formalism that we will use. For our purposes, boundary states live in the AP sector of the theory. As they preserve the conformal symmetry, and this includes $(-1)^F$, they in fact belong to the AP-even sector. This Hilbert space decomposes into a direct sum of Verma modules
\[ \colmath{\apeven}{\mathcal{H}_\text{AP}^\text{even}} = \bigoplus_{i,j \in \KT} M_{ij} \overline{\mathcal{V}_i} \otimes \mathcal{V}_j \]
where the multiplicities $M_{ij}$ can be read off from the partition function tables listed in Section~\ref{sec:review}. To preserve the conformal symmetry, boundary states can only contain states coming from terms with $i = j$, so must lie in the subspace
\[ \colmath{\apeven}{\mathcal{H}_\text{AP}^\text{even} \big|_\text{diagonal}} = \bigoplus_{i \in \KT} M_{ii} \overline{\mathcal{V}_i} \otimes \mathcal{V}_i = \bigoplus_{\substack{i \in \KT \\ M_{ii} = 1}} \overline{\mathcal{V}_i} \otimes \mathcal{V}_i \]
where in the last step, we've taken advantage of the fact that $M_{ii} = 0$ or 1 to simplify our answer. In each $\overline{\mathcal{V}_i} \otimes \mathcal{V}_i$, there is a unique conformally-invariant state $\kket{i}$ whose inner product with the vacuum state is 1, called an Ishibashi state \cite{ishibashi}. Any boundary state must therefore take the form
\[ \ket{a} = \sum_{\substack{i \in \KT \\ M_{ii} = 1}} a_i \kket{i} \]
for some set of coefficients $a_i$, which we assume to be real.\footnote{The phases of the $\kket{i}$ are actually ambiguous. We assume that these choices of phase have been fixed to make the $a_i$ real. As we will see, this can always be consistently done.}

When we impose boundary states $\bra{a}$ and $\ket{b}$ on an interval, the partition function that counts states on the interval is
\[ \text{Tr}_{\mathcal{H}_{ab}}(q^{(L/\pi)H}) = \sum_{i \in \KT} \chi_i(\tau) \sum_{\substack{j \in \KT \\ M_{jj} = 1}} \mathcal{S}_{ij} a_j b_j \]
where $q = e^{2 \pi i \tau}$ and $L$ is the length of the interval. We demand that the coefficients
\begin{equation} n_{ab}^i = \sum_j \mathcal{S}_{ij} a_j b_j \label{eqn:cardycond}\end{equation}
occurring in this expression are either positive integers or $\sqrt{2}$ times positive integers. This is a weakened version of Cardy's condition \cite{cardy} that allows for the presence of unpaired Majorana modes. Finally, we will look for a basis of such solutions, known as fundamental boundary states \cite{minimalstates}, defined by imposing the additional requirement that
\[ n_{ab}^0 = \delta_{ab} \]
Here $i = 0 \in \KT$ denotes the identity module, $(r, s) = (1, 1)$. All other solutions can then be expressed as a linear combination of the fundamental ones with nonnegative-integer coefficients.

Our goal in what follows will be to simply write down a complete set of fundamental boundary states consistent with all of the above properties. We do this for the infinite series and the exceptionals in turn.

\subsection{Infinite Series}

For these models, the quickest way to get the boundary states is to start with those of the underlying diagonal bosonic minimal model, and project onto their even part under the model's unique $\mathbb{Z}_2$ global symmetry. After discarding duplicates, and suitably adjusting the normalisation, we will have our answer.

We begin by recalling that the boundary states of the $m$th diagonal bosonic minimal model are given by the Cardy ansatz \cite{cardy}
\begin{equation} \ket{i}_\text{B} = \sum_{j \in \KT} \frac{\mathcal{S}_{ij}}{\sqrt{\mathcal{S}_{0j}}} \kket{j} \label{eqn:cardyansatz}\end{equation}
where $i \in \KT$ labels the states, and the `B' stands for bosonic. As it stands, these are \emph{not} valid boundary states of the fermionic theory, because the sum includes all states $\kket{j}$, whereas it should only include those with $M_{jj} = 1$. Looking back at the partition function tables in Section~\ref{sec:review}, we see the latter condition is equivalent to
\[ (m+1)r + ms = 1 \mod 2 \]
where $j = (r,s)$. But this is the same as the condition for $\kket{j}$ to be even under the $\mathbb{Z}_2$ global symmetry of the bosonic model \cite{ruelle}. Indeed, this symmetry acts as
\[ U \kket{j} = (-1)^{(m+1)r + ms + 1} \kket{j} \]
It follows that if we simply define our fermionic boundary states to be the $\mathbb{Z}_2$-even projections of the bosonic ones, then the undesirable states $\kket{j}$ disappear from the sum, and we are left with
\begin{equation} \ket{i}_F \; \coloneqq \; \frac{1 + U}{2} \, \ket{i}_B \; = \; \sum_{\substack{j \in \KT \\ M_{jj} = 1}} \frac{\mathcal{S}_{ij}}{\sqrt{\mathcal{S}_{0j}}} \kket{j} \label{eqn:fstate}\end{equation}
which, for each $i \in \KT$, defines a valid fermionic boundary state.

\newpage

Next, we discard duplicates. Currently the states $\ket{i}_F$ are overcomplete. We wish to restrict the range of $i$ to eliminate this redundancy. This can be done by considering the action of the $\mathbb{Z}_2$ symmetry on the bosonic boundary states \eqref{eqn:cardyansatz}, which is
\[ U \ket{i}_B = \ket{i'}_B \]
Here $i \rightarrow i'$ is the involution of the Kac Table defined by
\begin{equation} (r, s) \rightarrow (r, s)' = (m - r, s) \label{eqn:involution}\end{equation}
which corresponds to fusion with the special primary $(m - 1, 1)$. From \eqref{eqn:fstate}, we see that $\ket{i}_F = \ket{i'}_F$. It follows that we can eliminate the redundancy by restricting $i$ to lie in a set of representatives for the equivalence classes of $\KT / (i \sim i')$. One possible choice of a set of representatives, which we will use, is to take the bottom-left quadrant of the Kac Table, defined by the inequalities
\[ r \leq m / 2 \quad \text{and} \quad s \leq (m + 1) / 2 \]
This is because the involution \eqref{eqn:involution} acts as a horizontal reflection of the Kac Table, while the Kac Table is itself defined modulo a combined horizontal + vertical reflection.

Finally we adjust the normalisation to ensure $n_{ab}^0 = \delta_{ab}$ between all pairs of states. To do this it will first prove useful to write the boundary states in the form
\[ \ket{i}_F = \frac{\ket{i}_B + \ket{i'}_B}{2} \]
This makes it easy to calculate the coefficients $n_{ij}^k$ between the fermionic boundary states ${}_F \bra{i}$ and $\ket{j}_F$ using known results for the bosonic boundary states \cite{cardy}. We find
\[ n_{ij}^k = \frac{\mathcal{N}_{ijk} + \mathcal{N}_{i'jk} + \mathcal{N}_{ij'k} + \mathcal{N}_{i'j'k}}{4} = \frac{\mathcal{N}_{ijk} + \mathcal{N}_{ij'k}}{2} \]
From this we can extract the multiplicity of the identity module,
\[ n_{ij}^0 = \delta_{ij} \begin{cases} 1 & i = i' \\ 1/2 & i \neq i' \end{cases} \]
This is telling us that in order to achieve $n_{ij}^0 = \delta_{ij}$, we must rescale the boundary states corresponding to the second case by $\sqrt{2}$. After performing this rescaling, we arrive at our final result
\[ \ket{i}_F = \begin{cases} \ket{i}_B & i = i' \\ \frac{\ket{i}_B + \ket{i'}_B}{\sqrt{2}} & i \neq i' \end{cases} \]
Not surprisingly, the two cases will turn out to correspond to the two classes of boundary states. We also take the opportunity to remark that the two cases have a very simple graphical interpretation: the first corresponds to the border of the quadrant, the second to the interior.

We are now in a position where we can list the two classes of boundary states, and demonstrate the consistency of their interval partition functions.
\begin{itemize}
\item Class 1 consists of the points $(r, s)$ in the interior of the bottom-left quadrant of the Kac Table, defined by the inequalities $1 \leq r < m/2$ and $1 \leq s < (m + 1) / 2$. The states take the form
\[ \ket{(r,s)}_F = \sqrt{2} \hspace{-0.3em} \sum_{\substack{1 \leq s' \leq r' < m \\ (m + 1)r' + ms' \text{ odd}}} \hspace{-0.3em} \frac{\mathcal{S}_{(r,s),(r',s')}}{\sqrt{\mathcal{S}_{(1,1),(r',s')}}} \, \kket{(r',s')} \]
\item Class 2 consists of the points $(r, s)$ on the border of the bottom-left quadrant of the Kac Table, defined by the inequalities $1 \leq r \leq m/2$ and $1 \leq s \leq (m + 1) / 2$ with one of the upper bounds saturated. The states take an almost identical form, differing only in normalisation:
\[ \ket{(r,s)}_F = \sum_{\substack{1 \leq s' \leq r' < m \\ (m + 1)r' + ms' \text{ odd}}} \hspace{-0.3em} \frac{\mathcal{S}_{(r,s),(r',s')}}{\sqrt{\mathcal{S}_{(1,1),(r',s')}}} \, \kket{(r',s')} \]
\end{itemize}
On an interval with boundary states $i$ and $j$, the answer for the coefficients $n_{ij}^k$ depends on the classes of $i$ and $j$. There are three possible combinations to consider. From our earlier results, they are
\begin{align*}
\text{1--1:} \qquad n_{ij}^k &= \mathcal{N}_{ijk} + \mathcal{N}_{ijk'} \\
\text{1--2 or 2--1:} \qquad n_{ij}^k &= \sqrt{2} \, \mathcal{N}_{ijk} \\
\text{2--2:} \qquad n_{ij}^k &= \mathcal{N}_{ijk}
\end{align*}
As promised, we see that since the fusion numbers $\mathcal{N}_{ijk}$ are nonnegative integers (in fact 0 or 1), these answers have the desired integrality or $\sqrt{2}$-integrality properties.

\subsection{Exceptionals}

For the exceptional models, instead of starting from the underlying bosonic model, it is perhaps simplest to write down the boundary states directly. We do this by choosing a pair of seed boundary states and applying fusion to generate the rest. Most of the details will be the same between the $m = 11$ and the $m = 12$ models, so we show the details only for the $m = 11$ model.

We start by consulting the table in Section~\ref{sec:review} to see when $M_{ii} = 1$. We see that this is true when $i = (r, s)$ with $s = 1, 5, 7, 11$ and $r$ odd. It follows that the admissible boundary states of the fermionic model must lie in the span
\begin{equation} \ket{a} \in \text{span} \Big\{ \kket{(r,1)}, \, \kket{(r,5)}, \, \kket{(r,7)}, \, \kket{(r,11)} : r \text{ odd} \Big\} \label{eqn:exc11span}\end{equation}
Our first goal will be to write down the simplest consistent boundary state obeying this property. To do this, we assume that we already know in advance what the coefficients $n_{aa}^i$ will be. We can then reverse-engineer a boundary state that does the job:
\[ a_i = \sqrt{\mathcal{S}_{ij} n_{aa}^j} \]
For most choices of the coefficients $n_{aa}^i$, the resulting boundary state $\ket{a}$ will not be allowed, due to containing extra states $\kket{i}$ not in the span \eqref{eqn:exc11span}. We would like to make the simplest choice of $n_{aa}^i$ such that it is allowed. Our claim is that this choice is
\[ n_{aa}^i = \delta_{i,(1,1)} + \delta_{i,(1,5)} + \delta_{i,(1,7)} + \delta_{i,(1,11)} \]

\paragraph{Proof}
To demonstrate the above assertion, we need to show that
\[ \mathcal{S}_{(r,s),(1,1)} + \mathcal{S}_{(r,s),(1,5)} + \mathcal{S}_{(r,s),(1,7)} + \mathcal{S}_{(r,s),(1,11)} = 0 \quad \text{for } s \not\in \{1,5,7,11\} \]
For this we invoke the explicit formula for the modular $\mathcal{S}$-matrix
\[ \mathcal{S}_{(r,s),(r',s')} = \sqrt{\frac{8}{m(m+1)}} \, \sin\!\left( \frac{\pi r r'}{m} \right) \sin\!\left( \frac{\pi s s'}{m+1} \right) (-1)^{(r + s) (r' + s')} \]
and well as the trigonometric identity
\[ \sum_{s'=1,5,7,11} \sin\!\left( \frac{\pi s s'}{12} \right) = \begin{cases} \sqrt{6} & s \in \{1,5,7,11\} \\ 0 & \text{otherwise} \end{cases} \]
Putting these together gives the promised result. \hfill $\square$

So far we have obtained a single consistent seed state $\ket{a}$ with coefficients
\[ a_i = \sqrt{\mathcal{S}_{i,(1,1)} + \mathcal{S}_{i,(1,5)} + \mathcal{S}_{i,(1,7)} + \mathcal{S}_{i,(1,11)}} \]
Next we cook up a second seed state $\ket{b}$ by making a different set of sign choices in the coefficients. We choose
\[ b_{(r,s)} = a_{(r,s)} \begin{cases} +1 & s = 1, 7 \\ -1 & s = 5, 11 \end{cases} \]
where we recall when we discuss the coefficients of a fermionic boundary state, $r$ is odd and $s = 1,5,7,11$. Our next claim is that the state $\ket{b}$ gives rise to the interval partition functions
\begin{equation} n_{bb}^i = n_{aa}^i \quad \text{and} \quad n_{ab}^i = \sqrt{2} \, ( \delta_{i,(1,4)} + \delta_{i,(1,8)} ) \label{eqn:excproofb}\end{equation}
This is telling us that the two seed states $\ket{a}$, $\ket{b}$ lie in different SPT classes, but are otherwise consistent.

\paragraph{Proof}
The first equation in \eqref{eqn:excproofb} is trivial since $a_i$ and $b_i$ differ only by signs. The computation of $n_{ab}^i$ is somewhat trickier. This time the key fact we need is
\[ (-1)^{[s = 5,11]} \sum_{s'=1,5,7,11} \sin\!\left( \frac{\pi s s'}{12} \right) = \sqrt{2} \, \sum_{s'=4,8} \sin\!\left( \frac{\pi s s'}{12} \right) \]
Using the above identity, we compute
\[ a_{(r,s)} b_{(r,s)} = (-1)^{[s = 5,11]} \sum_{s'=1,5,7,11} \mathcal{S}_{(r,s),(1,s')} = \sqrt{2} \, \sum_{s'=4,8} \mathcal{S}_{(r,s),(1,s')} \]
By \eqref{eqn:cardycond} the claimed result for $n_{ab}^i$ immediately follows. \hfill $\square$

With the two seed states in hand, all remaining boundary states can now be generated by fusion \cite{affleck, fuchs}. This is a recipe which takes as input a boundary state, which for us will be either $\ket{a}$ or $\ket{b}$, as well as a primary operator $i \in \KT$, and produces a new consistent boundary state. The recipe for the fused boundary states is
\[ \ket{a; i} = \sum_{\substack{j \in \KT \\ M_{jj} = 1}} a_j \, \frac{\mathcal{S}_{ij}}{\mathcal{S}_{0j}} \, \kket{j} \qquad \ket{b; i} = \sum_{\substack{j \in \KT \\ M_{jj} = 1}} b_j \, \frac{\mathcal{S}_{ij}}{\mathcal{S}_{0j}} \, \kket{j} \]
which reduce back to the seed states when $i = 0$. Our final claim is that to obtain a complete basis of fundamental boundary states, we should let $i$ range over
\[ i = (r, s) \quad \text{where} \quad r = 1,3,5,7,9 \text{ and } s = 1, 2 \]
This assertion will become evident in a moment when we list the interval partition functions between all pairs of boundary states. For now, we simply note that the counting works out: the above family contains $2 \cdot 5 \cdot 2 = 20$ states, which coincides with the number of allowed Ishibashi states $\kket{j}$, as these correspond to $j = (r, s)$ with $r = 1,3,5,7,9$ and $s = 1,5,7,11$.

We are now in a position to list the states for the $m = 11$ exceptional model, and their interval partition functions:
\begin{itemize}
\item Both classes are labelled by pairs $(r, s)$ with $r = 1,3,5,7,9$ and $s = 1,2$. Regarded as elements of the Kac Table, these are all distinct elements. The two classes of states, after some algebra, are
\begin{align*}
\ket{(r,s)}_1 &= \sum_{\substack{r' = 1,3,5,7,9 \\ s' = 1,5,7,11}} \left\{ \! \begin{smallmatrix*}[l] \alpha &:\; s'=1 \\ \beta &:\; s'=5 \\ \beta &:\; s'=7 \\ \alpha &:\; s'=11 \end{smallmatrix*} \! \right\} \frac{\mathcal{S}_{(r,s),(r',s')}}{\sqrt{\mathcal{S}_{(1,1),(r',s')}}} \, \kket{(r',s')} \\
\ket{(r,s)}_2 &= \sum_{\substack{r' = 1,3,5,7,9 \\ s' = 1,5,7,11}} \left\{ \! \begin{smallmatrix*}[l] \alpha &:\; s'=1 \\ -\beta &:\; s'=5 \\ \beta &:\; s'=7 \\ -\alpha &:\; s'=11 \end{smallmatrix*} \! \right\} \frac{\mathcal{S}_{(r,s),(r',s')}}{\sqrt{\mathcal{S}_{(1,1),(r',s')}}} \, \kket{(r',s')}
\end{align*}
where $\alpha = \sqrt{2(3+\sqrt{3})}$ and $\beta = \sqrt{2(3-\sqrt{3})}$.
\item The interval partition functions are
\begin{align*}
\text{1--1 or 2--2:} \qquad n_{ij}^k &= \mathcal{N}_{ij}^x \, [\, \mathcal{N}_{x,(1,1)}^k + \mathcal{N}_{x,(1,5)}^k + \mathcal{N}_{x,(1,7)}^k + \mathcal{N}_{x,(1,11)}^k \,] \\
\text{1--2 or 2--1:} \qquad n_{ij}^k &= \sqrt{2} \; \mathcal{N}_{ij}^x \, [\, \mathcal{N}_{x,(1,4)}^k + \mathcal{N}_{x,(1,8)}^k \,]
\end{align*}
where the appearance of the fusion numbers is from Verlinde's formula \cite{verlinde}.
\end{itemize}
Before we go on, we return to an earlier point and verify the completeness of the states. Using the above formulas, the multiplicity of the identity module is
\begin{align*}
\text{1--1 or 2--2:} \qquad n_{ij}^0 &= \delta_{ij} + \mathcal{N}_{ij}^{(1,5)} + \mathcal{N}_{ij}^{(1,7)} + \mathcal{N}_{ij}^{(1,11)} \\
\text{1--2 or 2--1:} \qquad n_{ij}^0 &= \sqrt{2} \, [\, \mathcal{N}_{ij}^{(1,4)} + \mathcal{N}_{ij}^{(1,8)} \,]
\end{align*}
One can show that for the range of values of $i$ and $j$ allowed, all the fusion numbers in the above expression vanish identically, and only the $\delta_{ij}$ term survives. This shows that $n_{ab}^0 = \delta_{ab}$ between all pairs of states $a$ and $b$, establishing completeness.

We now briefly turn to the $m = 12$ exceptional. Here, the results are virtually identical, except with the roles of $r$ and $s$ swapped around:
\begin{itemize}
\item Both classes are labelled by pairs $(r, s)$ with $r = 1,2$ and $s = 1,3,5,7,9,11$. Regarded as elements of the Kac Table, these are again distinct. The states are
\begin{align*}
\ket{(r,s)}_1 &= \sum_{\substack{r' = 1,5,7,11 \\ s' = 1,3,5,7,9,11}} \left\{ \! \begin{smallmatrix*}[l] \alpha &:\; r'=1 \\ \beta &:\; r'=5 \\ \beta &:\; r'=7 \\ \alpha &:\; r'=11 \end{smallmatrix*} \! \right\} \frac{\mathcal{S}_{(r,s),(r',s')}}{\sqrt{\mathcal{S}_{(1,1),(r',s')}}} \, \kket{(r',s')} \\
\ket{(r,s)}_2 &= \sum_{\substack{r' = 1,5,7,11 \\ s' = 1,3,5,7,9,11}} \left\{ \! \begin{smallmatrix*}[l] \alpha &:\; r'=1 \\ -\beta &:\; r'=5 \\ \beta &:\; r'=7 \\ -\alpha &:\; r'=11 \end{smallmatrix*} \! \right\} \frac{\mathcal{S}_{(r,s),(r',s')}}{\sqrt{\mathcal{S}_{(1,1),(r',s')}}} \, \kket{(r',s')}
\end{align*}
with the same constants $\alpha$ and $\beta$ as before.
\item The interval partition functions are
\begin{align*}
\text{1--1 or 2--2:} \qquad n_{ij}^k &= \mathcal{N}_{ij}^x \, [\, \mathcal{N}_{x,(1,1)}^k + \mathcal{N}_{x,(5,1)}^k + \mathcal{N}_{x,(7,1)}^k + \mathcal{N}_{x,(11,1)}^k \,] \\
\text{1--2 or 2--1:} \qquad n_{ij}^k &= \sqrt{2} \; \mathcal{N}_{ij}^x \, [\, \mathcal{N}_{x,(4,1)}^k + \mathcal{N}_{x,(8,1)}^k \,]
\end{align*}
\end{itemize}

\section{Anomalous Symmetries} \label{sec:syms}

Here we turn to the second main goal of the paper, which is to list all $\mathbb{Z}_2$ global symmetries of the fermionic minimal models -- including potentially those with an anomaly. The motivation for performing this task comes from looking at the boundary states of the four special models for which the classes have equal sizes. In Section~\ref{sec:states}, we found these states to be
\begin{itemize}
\item Infinite series, $m = 3$:
\[ \kket{0} \pm \kket{\tfrac{1}{2}} \]
(Here and in the next example, we label Ishibashi states by conformal dimensions rather than elements of the Kac Table.)
\item Infinite series, $m = 4$:
\begin{align*}
&\tfrac{(5-\sqrt{5})^{1/4}}{10^{1/4}} \Big( \kket{0} \pm \kket{\tfrac{3}{2}} \Big) + \tfrac{(5+\sqrt{5})^{1/4}}{10^{1/4}} \Big( \kket{\tfrac{3}{5}} \pm \kket{\tfrac{1}{10}} \Big) \\
&\tfrac{(5+\sqrt{5})^{3/4}}{2^{3/4}\sqrt{5}} \Big( \kket{0} \pm \kket{\tfrac{3}{2}} \Big) - \tfrac{(5-\sqrt{5})^{3/4}}{2^{3/4}\sqrt{5}} \Big( \kket{\tfrac{3}{5}} \pm \kket{\tfrac{1}{10}} \Big)
\end{align*}
\item Exceptional, $m = 11$:
\[ \sum_{\substack{r' = 1\dots9,\text{odd} \\ s' = 1,5,7,11}} \left\{ \! \begin{smallmatrix*}[l] \alpha &:\; s'=1 \\ \pm\beta &:\; s'=5 \\ \beta &:\; s'=7 \\ \pm\alpha &:\; s'=11 \end{smallmatrix*} \! \right\} \frac{\mathcal{S}_{(r,s),(r',s')}}{\sqrt{\mathcal{S}_{(1,1),(r',s')}}} \, \kket{(r',s')} \]
\item Exceptional, $m = 12$:
\[ \sum_{\substack{r' = 1,5,7,11 \\ s' = 1\dots11,\text{odd}}} \left\{ \! \begin{smallmatrix*}[l] \alpha &:\; r'=1 \\ \pm\beta &:\; r'=5 \\ \beta &:\; r'=7 \\ \pm\alpha &:\; r'=11 \end{smallmatrix*} \! \right\} \frac{\mathcal{S}_{(r,s),(r',s')}}{\sqrt{\mathcal{S}_{(1,1),(r',s')}}} \, \kket{(r',s')} \]
\end{itemize}
where, in all cases, the upper choice of sign for the $\pm$ corresponds to the first class and the lower choice to the second. The above formulas give rise to an important observation: the two classes differ only by flipping the sign of a certain subset of the Ishibashi states. This strongly suggests that the classes are related by the action of a $\mathbb{Z}_2$ global symmetry. Our goal in this section will be to show that these models do indeed have a unique $\mathbb{Z}_2$ symmetry that exchanges the classes, while the remaining models have no such symmetry.

\subsubsection*{Anomalous $\mathbb{Z}_2$ Symmetries}

We start by reviewing some basic formalism about $\mathbb{Z}_2$ symmetries. Our perspective is similar to that described in \cite{znbootstrap}. The Hilbert space in the AP sector is
\[ \mathcal{H}_\text{AP} = \colmath{\apeven}{\mathcal{H}_\text{AP}^\text{even}} \oplus \colmath{\apodd}{\mathcal{H}_\text{AP}^\text{odd}} = \bigoplus_{i,j \in \KT} N_{ij} \overline{\mathcal{V}_i} \otimes \mathcal{V}_j = \bigoplus_{\substack{i,j \in \KT \\ N_{ij} = 1}} \overline{\mathcal{V}_i} \otimes \mathcal{V}_j \]
where the multiplicities $N_{ij}$ can be read off from the partition function tables in Section~\ref{sec:review}. Again, these only take the values $N_{ij} = 0, 1$, allowing us to make the final step. A $\mathbb{Z}_2$ symmetry must act on each $\overline{\mathcal{V}_i} \otimes \mathcal{V}_j$ by a sign, which we will denote as $s_{ij} = \pm 1$. These signs determine the partition function on a background with an AP-AP spin structure and a symmetry defect along the space direction:
\[ \mathcal{Z} \Big[ \tau; \, \spinh{AP}{AP} \; \Big] \; = \; \sum_{\substack{i,j \in \KT \\ N_{ij} = 1}} s_{ij} \, \overline{\chi_i(\tau)} \, \chi_j(\tau) \]
Knowledge of this one partition function then allows further partition functions to be determined. This is because the partition functions on various backgrounds are related among each other by acting on $\tau$ with modular transformations $\mathcal{S}(\tau) = -1 / \tau$ and $\mathcal{T}(\tau) = \tau + 1$. The relationships we need are encoded by the diagram
\begin{equation*}
\begin{tikzcd}
{\mathcal{Z}\Big[\tau;\spinh{AP}{AP}\;\Big]} \arrow[dash, d, "\mathcal{T}"] \arrow[dash, r, "\mathcal{S}"] &
{\mathcal{Z}\Big[\tau;\spinv{AP}{AP}\;\Big]} \arrow[dash, r, "\mathcal{T}"] &
{\mathcal{Z}\Big[\tau;\spind{P}{AP}\;\Big]} \arrow[dash, d, "\mathcal{S}"] \\
{\mathcal{Z}\Big[\tau;\spinh{P}{AP}\;\Big]} \arrow[dash, r, "\mathcal{S}"] &
{\mathcal{Z}\Big[\tau;\spinv{AP}{P}\;\Big]} \arrow[dash, r, "\mathcal{T}"] &
{\mathcal{Z}\Big[\tau;\spind{AP}{P}\;\Big]}
\end{tikzcd}
\end{equation*}
which allows all partition functions to be determined from the one on the top-left. A consistent $\mathbb{Z}_2$ symmetry is one for which all these partition functions admit a sensible expansion into Virasoro characters. For most of the backgrounds,
\[ \spinh{AP}{AP} \qquad \spinh{P}{AP} \qquad \spind{P}{AP} \qquad \spind{AP}{P} \]
this means a sum weighted by integers. However for the two special backgrounds
\[ \spinv{AP}{AP} \qquad \spinv{AP}{P} \]
the partition function is untwisted by any symmetries, and so we impose the stronger constraint that the weights be \emph{nonnegative} integers.

Actually, what we have described is not quite correct for anomalous symmetries. In this case we must weaken the above requirements in several ways. First, the diagram need only hold projectively, meaning some of the relationships expressed by the edges are violated by a phase. Second, the partition functions themselves are ambiguously defined up to a phase. By adjusting these phases if necessary, which amounts to making a choice of gauge for the diagram, we can always cast the diagram into the form
\begin{equation}
\begin{tikzcd}[column sep=large]
{\mathcal{Z}\Big[\tau;\spinh{AP}{AP}\;\Big]} \arrow[dash, d, "\mathcal{T}"] \arrow[dash, r, "\mathcal{S}"] &
{\mathcal{Z}\Big[\tau;\spinv{AP}{AP}\;\Big]} \arrow[dash, r, "\mathcal{T} \, = \, e^{+\pi i k / 8}"] &
{\mathcal{Z}\Big[\tau;\spind{P}{AP}\;\Big]} \arrow[dash, d, "\mathcal{S}"] \\
{\mathcal{Z}\Big[\tau;\spinh{P}{AP}\;\Big]} \arrow[dash, r, "\mathcal{S}"] &
{\mathcal{Z}\Big[\tau;\spinv{AP}{P}\;\Big]} \arrow[dash, r, "\mathcal{T} \, = \, e^{-\pi i k / 8}"] &
{\mathcal{Z}\Big[\tau;\spind{AP}{P}\;\Big]}
\end{tikzcd}
\label{eqn:orbitdiagramanom}
\end{equation}
where the phase violations are expressed by the notation $\mathcal{Z}[\tau; A] \xnoarrow{\mathcal{T} = e^{i\theta}} \mathcal{Z}[\tau; B]$, which means that $\mathcal{Z}[\tau + 1; A] = e^{i\theta} \mathcal{Z}[\tau; B]$ and vice-versa.%
\footnote{The argument for why this pattern of phases is universal is that the holonomy of a closed loop in the diagram is the value of $\exp(-\frac{i \pi}{2} \eta(\mathcal{D}))^k$ on a suitable mapping torus, where $\mathcal{D}$ is a certain 3d Dirac operator. Then because the phases are universal, they can be read off from the Majorana fermion. For more details, see \cite{wittenphases}, and for this particular example, also \cite{anomalyinterplay}. Another, more concrete way to see the pattern of phases is to use the identities $\mathcal{S}^2 = (\mathcal{ST})^3 = 1$, and the fact that the top-left partition function is invariant under $\mathcal{T}^2$ -- though with this approach the quantisation of $k$ is less obvious.}
The integer $k$ is the strength of the anomaly, and is valued mod 8.%
\footnote{Although $k$ appears as the exponent of a 16th root of unity, the shift $k \rightarrow k + 8$ is a gauge transformation, so $k$ is valued mod 8 not mod 16. We could have made $k$ manifestly mod-8 valued by making a different gauge choice. But the one we have chosen is more convenient in the long run.}

To use the diagram, we start with the top-left partition function, and determine the other partition functions and the integer $k$ by insisting that the diagram commutes. If no such integer $k$ can be found, we do not have a consistent symmetry. Otherwise we demand that the partition functions are sensible, as before. But because we may have needed to adjust their phases to gauge-fix the diagram, it only makes sense to demand that they have a sensible expansion into characters up to an overall phase.

There is one final consequence of the anomaly, and that is the possible appearance of an unpaired Majorana mode in the frustrated sectors. This means that for the backgrounds
\[ \spinv{AP}{AP} \qquad \spind{P}{AP} \qquad \spinv{AP}{P} \qquad \spind{AP}{P} \]
with a symmetry defect that wraps vertically, we should allow a possible overall factor of $\sqrt{2}$ in the partition function. We previously saw an example of this phenomenon in Section~\ref{sec:int:maj}.

\subsubsection*{Solving the Constraints}

We turn now to the task of solving the above constraints. To do this, we will recast a subset of the constraints as a set of matrix equations, solve them, and then check that the resulting solutions satisfy the remaining constraints.

We begin by deriving the set of matrix equations. The partition function on $\sspinh{AP}{AP}$ is encoded by the matrix
\[ A_{ij} = N_{ij} s_{ij} \]
where we recall that $N_{ij}$ is a known matrix of 0s and 1s encoding the partition function on $\sspin{AP}{AP}$, and the $s_{ij}$ are a set of unknown signs -- unknown except for that of the identity operator, which we set to be $s_{11} = +1$. Meanwhile, the corresponding matrix for the $\sspinv{AP}{AP}$ partition function must take the form
\[ [\sqrt{2} \,] \, e^{i \theta} B_{ij} \]
where $B$ is an unknown matrix of nonnegative integers, the phase $\theta$ is arbitrary, and the $\sqrt{2}$ may or may not be present. Our first equation arises from the fact that $\sspinh{AP}{AP}$ and $\sspinv{AP}{AP}$ are related by an $\mathcal{S}$-transformation. This fact is expressed by
\[ \mathcal{S} A \, \mathcal{S} = [\sqrt{2} \,] \, e^{i \theta} B \]
Actually, since $\mathcal{S}$ is real, the phase $e^{i\theta}$ must equal $\pm 1$, so we obtain
\begin{equation} \mathcal{S} A \, \mathcal{S} = \pm [\sqrt{2} \,] B \label{eqn:srelation}\end{equation}
A second equation arises by considering a $\mathcal{T}^2$ transformation of $\sspinh{AP}{AP}$. Such a transformation preserves the background, but contributes an anomalous phase of $e^{2 \pi i k / 8}$. This fact is expressed by
\begin{equation} \mathcal{T}^{-2} B \, \mathcal{T}^2 = e^{2 \pi i k / 8} B \label{eqn:trelation}\end{equation}
Equations \eqref{eqn:srelation} and \eqref{eqn:trelation} will be all we need to determine the symmetries. We shall analyse them separately depending on whether the $\sqrt{2}$ is present or absent from \eqref{eqn:srelation}.

\subsection{The Case of No Majorana Mode}

If \eqref{eqn:srelation} contains no $\sqrt{2}$, then we can easily show that the only solutions are the trivial symmetry and fermion parity. To do this, we will make use of various Galois-theoretic results that were used extensively to solve the bosonic version of this problem \cite{ruelle}. As these results will play an important role both in this section and the next, we begin with a brief review of these ideas.

The entries of the modular $\mathcal{S}$-matrix belong to the cyclotomic field $\mathbb{Q}(\zeta_n)$ where $n = 2m(m+1)$, and $\zeta_n = e^{2 \pi i / n}$ is an $n$th root of unity. This field is acted on by Galois transformations $\sigma_h$, labelled by elements $h \in \mathbb{Z}_n^*$, via
\[ \sigma_h(\zeta_n) = \zeta_n^h \]
The action of $\sigma_h$ on the modular $\mathcal{S}$-matrix is \cite{costegannon}
\[ \sigma_h(\mathcal{S}_{ij}) = \epsilon_h(i) \, \mathcal{S}_{\sigma_h(i) j} \]
where $\epsilon_h(i)$ is a sign given by
\[ \epsilon_h(r, s) = \eta_h \, \epsilon_m(hr) \, \epsilon_{m+1}(hs) \]
while $\sigma_h$ is a permutation that will not be of interest to us. Here we have also introduced $\eta_h = \sigma_h(\sqrt{n}) / \sqrt{n}$, a computable but irrelevant sign, as well as another sign
\[ \epsilon_m(x) = \mathrm{sign} \sin \! \left( \frac{\pi x}{m} \right) \]
known as an $\mathfrak{su}(2)$ affine parity, defined for all $x \neq 0$ mod $m$.

We can use these facts to derive a useful constraint on $B_{ij}$. Starting from \eqref{eqn:srelation}, applying $\sigma_h$, and comparing the result back to \eqref{eqn:srelation} gives
\[ B_{ij} = \epsilon_h(i) \, \epsilon_h(j) \, B_{\sigma_h(i) \sigma_h(j)} \]
Crucially, all the entries of $B_{ij}$ are nonnegative. This means that if ever the product of signs $\epsilon_h(i) \epsilon_h(j)$ equals $-1$ for any $h$, then both sides must be zero. We learn that $B_{ij}$ obeys the \emph{parity rule}
\begin{equation} B_{ij} \neq 0 \quad \text{only if} \quad \epsilon_h(i) = \epsilon_h(j) \; \forall h \label{eqn:parityrule}\end{equation}
The set of all pairs $(i, j)$ obeying this condition was determined in \cite{ruelle}, and by Result~4 of that paper, they have conformal dimensions satisfying
\[ h_i - h_j \in \tfrac{1}{2} \mathbb{Z} \cup \tfrac{1}{3} \mathbb{Z} \cup \tfrac{1}{5} \mathbb{Z} \]
Now we recall \eqref{eqn:trelation}, whose $(i, j)$th component reads
\[ e^{4 \pi i (h_j - h_i)} B_{ij} = e^{2 \pi i k / 8} B_{ij} \]
The phase on the left hand side can never be a nontrivial power of an eight root of unity. We learn that $k = 0$, or in other words, that the symmetry is non-anomalous.

The existence of a non-anomalous symmetry is a powerful statement. It implies that if we perform a GSO projection, then the symmetry survives in the resulting bosonic theory \cite{spintqfts, statesum, web2d, orbifoldgroupoids}. Let us call the original symmetry $\alpha$, and denote by $\iota(\alpha)$ its image in the bosonic theory. Then $\iota(\alpha)$ commutes with $\iota((-1)^F)$. In the bosonic minimal models, the only such symmetry is $\iota((-1)^F)$ itself, hence $\iota(\alpha) = \iota((-1)^F)$, which in turn forces $\alpha = (-1)^F$. We therefore learn that the original symmetry must have been either trivial or fermion parity, as claimed.

We note that it would have also been possible to derive this conclusion directly from equations \eqref{eqn:srelation} and \eqref{eqn:trelation}, using technical arguments entirely parallel to those in \cite{ruelle}. However the above argument, similar to those used in \cite{lessonsramond}, is more transparent.

\subsection{The Case of a Majorana Mode} \label{sec:syms:anom}

In the more novel case that \eqref{eqn:srelation} contains a $\sqrt{2}$, we show that solutions exist only for $m = 3,4$ in the infinite series and $m = 11,12$ in the exceptionals.

As a first step we show that $m = 3, 4$ mod 8. Our starting point is \eqref{eqn:srelation}, from which we can immediately deduce
\[ \sqrt{2} \in \mathbb{Q}(\zeta_n) \]
This is true if and only if $n$ is a multiple of 8, which in turn requires
\[ m = 0,3 \text{ mod } 4 \]
Assuming from now on that this is the case, we can return to \eqref{eqn:srelation} and repeat our earlier derivation of the parity rule \eqref{eqn:parityrule} obeyed by $B_{ij}$. We find that it is modified by the presence of the $\sqrt{2}$, and now takes the form
\begin{equation} B_{ij} \neq 0 \quad \text{only if} \quad \epsilon_h(i) = f(h) \, \epsilon_h(j) \; \forall h \label{eqn:parityrulemod}\end{equation}
Here $f(h) = \sigma_h(\sqrt{2}) / \sqrt{2}$ is a sign which comes from the Galois transformation of the factor of $\sqrt{2}$, and takes the explicit form
\[ f(h) = \begin{cases} +1 & h = 1,7 \text{ mod } 8 \\ -1 & h = 3,5 \text{ mod } 8 \end{cases} \]
To solve condition \eqref{eqn:parityrulemod} for $(i, j)$, we write it in terms of affine parities as
\[ \epsilon_m(hr_i) \, \epsilon_{m+1}(hs_i) = f(h) \, \epsilon_m(hr_j) \, \epsilon_{m+1}(hs_j) \quad \forall h \in \mathbb{Z}_{2n}^* \]
Let us assume that $m = 0$ mod 4, since the case $m = 3$ mod 4 can be treated almost identically. Then the above condition is equivalent to
\begin{align*}
\epsilon_m(hr_i) &= f(h) \, \epsilon_m(hr_j) \quad \forall h \in \mathbb{Z}_{2m}^* \\
\epsilon_{m+1}(hs_i) &= \epsilon_{m+1}(hs_j) \quad \forall h \in \mathbb{Z}_{2(m+1)}^*
\end{align*}
The solutions to the second equation were determined in \cite{ruelle}, where Result 3 states that $s_i = s_j$ or $s_i = m + 1 - s_j$. Meanwhile, the solutions to the first equation are determined by the following conjecture:

\paragraph{Conjecture} Suppose $n$ is a multiple of 4, $1 \leq x, y \leq n - 1$, and $\epsilon_n(hx) = f(h) \, \epsilon_n(hy)$ for all $h \in \mathbb{Z}_{2n}^*$. Then the only solutions are
\begin{itemize}
\item $n = 4$ and $(x, y) = (1, 2)$,
\item $n = 8$ and $(x, y) = (1, 3)$,
\item $n = 12$ and $(x, y) = (1, 4), (2, 3), (4, 5)$,
\item $n = 24$ and $(x, y) = (1, 7), (5, 11)$,
\end{itemize}
and those that can be obtained from them by the solution-preserving transformations
\[ (n, x, y) \; \rightarrow \; (n, y, x),\, (n, n - x, y),\, (n, x, n - y),\, (kn, kx, ky) \]
It is computationally trivial to verify this conjecture up to $n = 1000$, and we expect, but have not proved, that it holds for all $n$.

We now know all solutions to \eqref{eqn:parityrulemod}. Without loss of generality, $s_i = s_j$. (This uses the equivalence $(r, s) \sim (m - r, m + 1 - s)$ of the Kac Table.) The values of $r_i$ and $r_j$ are given by
\begin{itemize}
\item if $m = 4a$, $(r_i, r_j) = (a, 2a)$,
\item if $m = 8a$, $(r_i, r_j) = (a, 3a)$,
\item if $m = 12a$, $(r_i, r_j) = (a, 4a), (2a, 3a), (4a, 5a)$,
\item if $m = 24a$, $(r_i, r_j) = (a, 7a), (5a, 11a)$,
\end{itemize}
and those related to them under $(r_i, r_j) \rightarrow (r_j, r_i), (m - r_i, r_j), (r_i, m - r_j)$.

Next, we use this result to compute the possible values of the phase $e^{4 \pi i (h_i - h_j)}$ among all solutions. Using the identity $2(h_i - h_j) = (m+1) \frac{r_i^2 - r_j^2}{2m}$ mod 1, we find the following contributions to the set of values taken by $e^{4 \pi i (h_i - h_j)}$:
\begin{itemize}
\item $m = 4a \; \rightarrow \; \zeta_8^{\pm a}$
\item $m = 8a \; \rightarrow \; (-1)^a$
\item $m = 12a \; \rightarrow \; \zeta_8^{\pm a}, \, \zeta_{24}^{\pm 7a}$
\item $m = 24a \; \rightarrow \; 1$
\end{itemize}
The payoff of these computations comes from considering the identity \eqref{eqn:trelation}, which in components takes the form
\[ e^{4 \pi i (h_j - h_i)} B_{ij} = e^{2 \pi i k / 8} B_{ij} \]
In view of the possible phases that can arise on the left hand side when $B_{ij} \neq 0$, it is easy to see that the anomaly $k$ must satisfy
\begin{align*}
m = 4 \text{ mod } 8 \quad &\implies \quad k = \text{odd} \\
m = 0 \text{ mod } 8 \quad &\implies \quad k = \text{even}
\end{align*}
All of the above analysis also goes through unchanged if instead $m = 3$ mod 4. Combining the results from the two cases, we conclude that
\begin{align*}
m = 3,4 \text{ mod } 8 \quad &\implies \quad k = \text{odd} \\
m = 0,7 \text{ mod } 8 \quad &\implies \quad k = \text{even}
\end{align*}
An even value of $k$ is incompatible with our assumption of a $\sqrt{2}$ in the frustrated partition function \cite{delmastro}. Indeed, if $k$ were even, then we could stack with an even number $8 - k$ of copies of the Majorana fermion, obtaining a system with no anomaly yet still with an unpaired Majorana zero in the frustrated sector. This gives a contradiction, since the presence of an unpaired zero mode is an exclusive feature of anomalous theories. The only way to avoid this contradiction is if
\[ m = 3, 4 \text{ mod } 8 \]
which is what we wanted to show.

While we have just ruled out $m = 0, 7$ mod 8 by a physical argument, one can also ask how the mathematical requirements \eqref{eqn:srelation} and \eqref{eqn:trelation} fail in this case. The answer is that \eqref{eqn:srelation} is generically violated, with $B_{ij}$ failing to be a nonnegative-integer matrix. The only exceptions are $m = 7, 8$, where there is a single solution for $B_{ij}$, but this then goes on to fail \eqref{eqn:trelation}.

To make further progress with the remaining cases, we deal with the infinite series and the exceptional models in turn.

\subsubsection{Infinite Series}

For the infinite series when $m = 3, 4$ mod 8, the matrix $N_{ij}$ takes the form
\begin{equation*}
N = \bigoplus_{\substack{i \in \KT/' \\ \xi_i = 1}}
\begin{blockarray}{ccc}
  & i & i' \\
  \begin{block}{r(cc)}
    i & 1 & 1 \\
    i' & 1 & 1 \\
  \end{block}
\end{blockarray}
\end{equation*}
This notation requires a little explanation. To each weight $i = (r, s) \in \KT$ there is an associated sign
\[ \xi_i = (-1)^{(m+1)r + ms + 1} \]
One can show that $\xi_i = \xi_{i'}$, and that $\xi_i = +1 \implies i \neq i'$. (Here $(r, s)' = (m - r, s)$ is the involution \eqref{eqn:involution} introduced in Section~\ref{sec:states}.) These facts are necessary to ensure the above block decomposition makes sense.

First we use the consistency conditions to constrain the form of $A_{ij}$, the matrix of unknown signs corresponding to $\sspinh{AP}{AP}$, using arguments analogous to \cite{ruelle}. Because $\mathcal{S} N \mathcal{S} = N$, which follows from $\mathcal{S}$-invariance of $\sspin{AP}{AP}$, $N_{ij}$ obeys the parity rule \eqref{eqn:parityrule}. Since $A_{ij} = N_{ij} s_{ij}$, so too does $A_{ij}$. By acting on \eqref{eqn:srelation} with a Galois transformation, comparing it back to \eqref{eqn:srelation} and invoking the previous fact, we find that $A_{ij}$ obeys a \emph{modified permutation rule}
\[ A_{\sigma_h(i) \sigma_h(j)} = f(h) A_{ij} \]
This states that the many unknown signs in $A_{ij}$ are in fact related. Rather than using the full power of this equation, though, we shall only use it for the special case $h = m(m+1) - 1$. In this case $\sigma_h(i) = i'$ and $f(h) = -1$, so we obtain
\[ A_{i'j'} = -A_{ij} \]
This states that the unknown signs within each $2 \times 2$ block of $A_{ij}$ are related, and that $A_{ij}$ takes the form
\begin{equation}
A = \bigoplus_{\substack{i \in \KT/' \\ \xi_i = 1}}
\begin{blockarray}{ccc}
  & i & i' \\
  \begin{block}{r(cc)}
    i & \epsilon_i & \eta_i \\
    i' & -\eta_i & -\epsilon_i \\
  \end{block}
\end{blockarray}
\label{amatrix}
\end{equation}
where $\epsilon_i$ and $\eta_i$ are a set of unknown signs associated to the elements of $\KT / '$, and the sign corresponding to the identity operator is $\epsilon_1 = +1$.

We are now in a position to rule out all but a finite number of models as having no symmetries. Recall that $B_{ij}$ obeys the modified parity rule \eqref{eqn:parityrulemod}. By the conjecture, this implies that unless $m \in \{ 3, 4, 11, 12 \}$, we have
\[ B_{1i} = B_{1i} = 0 \]
By \eqref{eqn:srelation} this tells us that $A$ annihilates the vector $\mathcal{S}_{1i}$ from both sides. Using \eqref{amatrix} and the fact $\mathcal{S}_{1i} = \mathcal{S}_{1i'} > 0$, we obtain the contradictory equations
\[ \epsilon_i = \pm \eta_i \]
showing immediately there are no solutions.

Finally we return to the special cases $m \in \{ 3, 4, 11, 12 \}$ that were exempt from the above no-go analysis. Listing the solutions for these cases is a purely finite problem, which can easily be done manually. We find the following results:
\begin{itemize}

\item \underline{$m = 3$} \hspace{1ex} The $m=3$ model has two symmetries. We specify them by writing out the partition function diagram \eqref{eqn:orbitdiagramanom} explicitly. For the first one, we have
\begin{equation*}
\begin{tikzcd}
{\overline{(\chi_0 + \chi_\frac{1}{2})}(\chi_0 - \chi_\frac{1}{2})} \arrow[dash, d] \arrow[dash, r] &
{\sqrt{2} \, \overline{(\chi_0 + \chi_\frac{1}{2})} \, \chi_\frac{1}{16}} \arrow[dash, r] &
{\sqrt{2} \, \overline{(\chi_0 - \chi_\frac{1}{2})} \, \chi_\frac{1}{16}} \arrow[dash, d] \\
{\overline{(\chi_0 - \chi_\frac{1}{2})}(\chi_0 + \chi_\frac{1}{2})} \arrow[dash, r] &
{\sqrt{2} \, \overline{\vphantom{(} \chi_\frac{1}{16}} (\chi_0 + \chi_\frac{1}{2})} \arrow[dash, r] &
{\sqrt{2} \, \overline{\vphantom{(} \chi_\frac{1}{16}} (\chi_0 - \chi_\frac{1}{2})}
\end{tikzcd}
\end{equation*}
with anomaly $k = 1$. The other symmetry is given by flipping the diagram upside down, an operation which corresponds to composing with $(-1)^F$, and has the opposite anomaly $k = -1$. These symmetries are of course simply left and right chiral fermion parity of the Majorana fermion.

\item \underline{$m = 4$} \hspace{1ex} The $m=4$ model has a similar structure, with two symmetries related by $(-1)^F$ and opposite anomalies. We shall therefore only give details of the first one. The partition function diagram \eqref{eqn:orbitdiagramanom} in this case takes the form
\begin{equation*}
\begin{tikzcd}[column sep=small, ampersand replacement=\&]
{\begin{alignedat}{2} &\overline{(\chi_0 + \chi_\frac{3}{2})}(\chi_0 - \chi_\frac{3}{2}) \\ {}+{} &\overline{(\chi_\frac{3}{5} + \chi_\frac{1}{10})}(\chi_\frac{3}{5} - \chi_\frac{1}{10}) \end{alignedat}} \arrow[dash, d] \arrow[dash, r] \&
{\begin{alignedat}{2} \sqrt{2} \, [ \, &\overline{(\chi_0 + \chi_\frac{3}{2})} \, \chi_\frac{7}{16} \\ {}+{} &\overline{(\chi_\frac{3}{5} + \chi_\frac{1}{10})} \, \chi_\frac{3}{80} \, ] \end{alignedat}} \arrow[dash, r] \&
{\begin{alignedat}{2} \sqrt{2} \, [ \, &\overline{(\chi_0 - \chi_\frac{3}{2})} \, \chi_\frac{7}{16} \\ {}+{} &\overline{(\chi_\frac{3}{5} - \chi_\frac{1}{10})} \, \chi_\frac{3}{80} \, ] \end{alignedat}} \arrow[dash, d] \\
{\begin{alignedat}{2} &\overline{(\chi_0 - \chi_\frac{3}{2})}(\chi_0 + \chi_\frac{3}{2}) \\ {}+{} &\overline{(\chi_\frac{3}{5} - \chi_\frac{1}{10})}(\chi_\frac{3}{5} + \chi_\frac{1}{10}) \end{alignedat}} \arrow[dash, r] \&
{\begin{alignedat}{2} \sqrt{2} \, [ \, &\overline{\chi_\frac{7}{16}} \, (\chi_0 + \chi_\frac{3}{2}) \\ {}+{} &\overline{\chi_\frac{3}{80}} \, (\chi_\frac{3}{5} + \chi_\frac{1}{10}) \, ] \end{alignedat}} \arrow[dash, r] \&
{\begin{alignedat}{2} \sqrt{2} \, [ \, &\overline{\chi_\frac{7}{16}} \, (\chi_0 - \chi_\frac{3}{2}) \\ {}+{} &\overline{\chi_\frac{3}{80}} \, (\chi_\frac{3}{5} - \chi_\frac{1}{10}) \, ] \end{alignedat}}
\end{tikzcd}
\end{equation*}
and the anomaly is $k = -1$.

\item \underline{$m = 11, 12$} \hspace{1ex} The models with $m = 11, 12$ turn out to have no symmetries.

\end{itemize}

\subsubsection{Exceptionals}

The above analysis can also be carried out for the two exceptional models at $m = 11, 12$. The results have the same structure as for $m = 3, 4$ earlier: there are two symmetries, related to each other by $(-1)^F$, with opposite anomalies. Below we list the symmetries for the two models. This time we will only list the partition functions on $\sspinh{AP}{AP}$ and $\sspinv{AP}{AP}$, as all others are related to these by the same pattern as in previous cases.

\begin{itemize}

\item \underline{$m = 11$} \hspace{1ex} For the $m = 11$ exceptional, we have
\begin{alignat*}{2}
\mathcal{Z}\Big[\tau;\spinh{AP}{AP}\;\Big] &={}& &\sum_{\mathclap{r = 1, \text{odd}}}^9 \overline{(\chi_{r,1} + \chi_{r,7} + \chi_{r,5} + \chi_{r,11})} (\chi_{r,1} + \chi_{r,7} - \chi_{r,5} - \chi_{r,11}) \\
\mathcal{Z}\Big[\tau;\spinv{AP}{AP}\;\Big] &={}& \sqrt{2} \, &\sum_{\mathclap{r = 1, \text{odd}}}^9 \overline{(\chi_{r,1} + \chi_{r,7} + \chi_{r,5} + \chi_{r,11})} (\chi_{r,4} + \chi_{r,8})
\end{alignat*}
with anomaly $k = -1$.

\item \underline{$m = 12$} \hspace{1ex} The $m = 12$ exceptional is similar, but with the roles of $r$ and $s$ reversed
\begin{alignat*}{2}
\mathcal{Z}\Big[\tau;\spinh{AP}{AP}\;\Big] &={}& &\sum_{\mathclap{s = 1, \text{odd}}}^{11} \overline{(\chi_{1,s} + \chi_{7,s} + \chi_{5,s} + \chi_{11,s})} (\chi_{1,s} + \chi_{7,s} - \chi_{5,s} - \chi_{11,s}) \\
\mathcal{Z}\Big[\tau;\spinv{AP}{AP}\;\Big] &={}& \sqrt{2} \, &\sum_{\mathclap{s = 1, \text{odd}}}^{11} \overline{(\chi_{1,s} + \chi_{7,s} + \chi_{5,s} + \chi_{11,s})} (\chi_{4,s} + \chi_{8,s})
\end{alignat*}
and anomaly $k = 1$.

\end{itemize}
Before going on, we pause to note that the earlier results for the infinite series can also be rewritten to look more like the results above. For $m = 3$, we have
\begin{alignat*}{2}
\mathcal{Z}\Big[\tau;\spinh{AP}{AP}\;\Big] &={}& &\sum_{\mathclap{r=1, \text{odd}}}^1 \overline{(\chi_{r,1} + \chi_{r,3})} (\chi_{r,1} - \chi_{r,3}) \\
\mathcal{Z}\Big[\tau;\spinv{AP}{AP}\;\Big] &={}& \sqrt{2} \, &\sum_{\mathclap{r=1, \text{odd}}}^1 \overline{(\chi_{r,1} + \chi_{r,3})} \, \chi_{r,2}
\end{alignat*}
while for $m = 4$ we have
\begin{alignat*}{2}
\mathcal{Z}\Big[\tau;\spinh{AP}{AP}\;\Big] &={}& &\sum_{\mathclap{s=1, \text{odd}}}^3 \overline{(\chi_{1,s} + \chi_{3,s})} (\chi_{1,s} - \chi_{3,s}) \\
\mathcal{Z}\Big[\tau;\spinv{AP}{AP}\;\Big] &={}& \sqrt{2} \, &\sum_{\mathclap{s=1, \text{odd}}}^3 \overline{(\chi_{1,s} + \chi_{3,s})} \, \chi_{2,s}
\end{alignat*}

\section{Discussion} \label{sec:discussion}

We conclude with some comments on our results. First we return to the earlier claim that the anomalous symmetries, where they exist, exchange the two classes of boundary states. This follows at a glance from the $\sspinh{AP}{AP}$ partition functions of Section~\ref{sec:syms:anom}. Indeed the coefficient of $|\chi_i|^2$ in this partition function determines the sign with which the symmetry acts on the Ishibashi state $\kket{i}$. For example, in the $m = 11$ exceptional, the charges of Ishibashi states are
\begin{center}
\begin{tabular}{cccc}
$\kket{(r,1)}$ & $\kket{(r,5)}$ & $\kket{(r,7)}$ & $\kket{(r,11)}$ \\
$+1$ & $-1$ & $+1$ & $-1$
\end{tabular}
\end{center}
These are precisely the charges needed to exchange the boundary states we wrote down in Section~\ref{sec:states}. The same conclusion also holds in the other models.

Independently of the details of any particular model, the fact the symmetry exchanges the classes can also be understood from the anomaly $k$ being odd. This follows by a stacking argument. Suppose a theory has a boundary state $\ket{a}$ and a symmetry $U$ with odd anomaly $k$. Then if we stack the theory with $-k$ mod 8 copies of the Majorana fermion, the resulting theory has a boundary state $\ket{a} \otimes \ket{+}^{-k}$ and a non-anomalous symmetry $U \otimes (-1)^{F_L}$. Acting on the state with the symmetry gives a new state $U \ket{a} \otimes \ket{-}^{-k}$, which must lie in the same class since the symmetry is non-anomalous. From Section~\ref{sec:int:maj}, we know that $\ket{+}^{-k}$ and $\ket{-}^{-k}$ are in different classes when $k$ is odd. Therefore so too must $\ket{a}$ and $U \ket{a}$, as claimed.

We would also like to return to a subtlety we felt was best left unaddressed in Section~\ref{sec:syms}, but are now in a position to close. A theory with a $\mathbb{Z}_2$ symmetry actually has six more partition functions that we did not consider. These fit into an orbit diagram, analogous to \eqref{eqn:orbitdiagramanom}, which looks like
\begin{equation*}
\begin{tikzcd}
{\mathcal{Z}\Big[\tau;\spinh{AP}{P}\;\Big]} \arrow[dash, loop, distance=4em, in=195, out=165, "\mathcal{T}"'] \arrow[dash, r, "\mathcal{S} = \zeta_8^{-k}"] &
{\mathcal{Z}\Big[\tau;\spinv{P}{AP}\;\Big]} \arrow[dash, r, "\mathcal{T} = \zeta_{16}^k"] &
{\mathcal{Z}\Big[\tau;\spind{AP}{AP}\;\Big]} \arrow[dash, loop, distance=4em, in=15, out=345, "\mathcal{S} = \zeta_8^{-k}"'] \\
{\mathcal{Z}\Big[\tau;\spinh{P}{P}\;\Big]} \arrow[dash, loop, distance=4em, in=195, out=165, "\mathcal{T}"'] \arrow[dash, r, "\mathcal{S} = \zeta_8^k"] &
{\mathcal{Z}\Big[\tau;\spinv{P}{P}\;\Big]} \arrow[dash, r, "\mathcal{T} = \zeta_{16}^{-k}"] &
{\mathcal{Z}\Big[\tau;\spind{P}{P}\;\Big]} \arrow[dash, loop, distance=4em, in=15, out=345, "\mathcal{S} = \zeta_8^k"']
\end{tikzcd}
\end{equation*}
In the models with an anomalous symmetry, all these partition functions are zero. This is because when $k = \pm 1$, the diagram violates the relations $S^2 = (ST)^3 = 1$ that must be satisfied by any minimal model partition functions, so there are no nonzero solutions. With these partition functions now in hand, one might ask why we did not demand the consistency of symmetry-projected traces, such as
\[ \frac{1}{2} \left( \mathcal{Z}\Big[\tau;\spinv{AP}{P}\;\Big] + \mathcal{Z}\Big[\tau;\spinv{P}{P}\;\Big] \right) \]
which counts bosonic states on a periodic circle frustrated by a symmetry defect. The answer becomes clear if we look at the Majorana fermion. Here, the answer would be
\[ \frac{1}{\sqrt{2}} \, \overline{(\chi_0 + \chi_{1/2})(\tau)} \, \chi_{1/16}(\tau) \]
which is inconsistent even for a fermionic theory. The issue is of course that with an unpaired Majorana mode, there is no separation into fermionic and bosonic states, and such symmetry-projected traces are meaningless. We conclude that there is no need to demand consistency of the symmetry-projected traces, and the constraints we have imposed on our symmetries are all that there are.

Next we attempt to shed some light on the observation that, from our results, the following properties of fermionic minimal models appear to be equivalent:
\[ \begin{matrix}\text{1. equal} \\ \text{class sizes}\end{matrix} \iff \begin{matrix}\text{2. existence of an} \\ \text{anomalous symmetry}\end{matrix} \iff \begin{matrix}\text{3. vanishing of the PP} \\ \text{sector partition function}\end{matrix} \]
Indeed, all three conditions are satisfied by $m = 3, 4$ from the infinite series and $m = 11, 12$ from the exceptionals. It is natural to ask whether the above superficial equivalences are in fact honest equivalences. Below we will outline some arguments that show that for some for them at least, the equivalences are indeed honest.

\begin{description}
\item[$\bm{2 \Rightarrow 1}$] ${}$ \\ As we have seen, when an anomalous symmetry is present it exchanges the two classes of boundary states. This trivially implies they have equal sizes.
\item[$\bm{3 \Rightarrow 1}$] ${}$ \\ Here we can argue the contrapositive as follows. SPT classes form an affine space: they can be compared, but there is no preferred choice of one phase as trivial. If however a model has boundary state classes of different sizes, then we seemingly have a way to distinguish one class over the other. But this is not a contradiction, as we should remember that for every fermionic minimal model listed in Section~\ref{sec:review}, there is another related by stacking with $(-1)^\text{Arf}$ \cite{spintqfts, statesum, web2d}, and whose boundary state class sizes are reversed. So it is acceptable for a theory to have different class sizes as long as $\mathcal{Z}$ and $\mathcal{Z} (-1)^\text{Arf}$ are different theories. The condition for this is that $\mathcal{Z}$ does not vanish in the PP sector.
\item[$\bm{2 \Rightarrow 3}$] ${}$ \\ This observation, that anomalies can force the vanishing of the Ramond sector, has been noted in various places in the literature. See for example \cite{delmastro}, where it was explained using the algebra obeyed by $(-1)^{F_L}$, $(-1)^{F_R}$ and $(-1)^F$.
\end{description}
\noindent For the above implications, we do not know of arguments in the other direction.

Finally we comment on our Conjecture in Section~\ref{sec:syms:anom}. In earlier work on the bosonic minimal models, the role of this conjecture was played by Result~14 of \cite{ruelle}. This result was proved by relating it to a theorem about the classification of simple factors of Jacobians of Fermat Curves \cite{aoki}. Given the close connection between the fermionic and bosonic minimal models, it is natural to ask if our conjecture has a similar interpretation in terms of Fermat curves. Whether or not it does remains an open question.

\section*{Acknowledgements}

The author thanks Shu-Heng Shao for the initial idea for this paper, David Tong for guidance and comments on the draft, Joe Davighi and Nakarin Lohitsiri for many discussions about anomalies, and Pietro Benetti Genolini, Avner Karasik and Carl Turner for other enlightening discussions. PBS acknowledges support from the Vice Chancellor's Award and David Tong's Simons Investigator Award.


\bibliographystyle{jhep}
\bibliography{refs}

\end{document}